\documentclass{svproc}
\usepackage[utf8]{inputenc}

\usepackage{url}

\usepackage{natbib}
\bibpunct{(}{)}{;}{a}{}{,}%
\usepackage{graphicx}
\usepackage{amsmath,amssymb}
\usepackage{multirow}
\usepackage{longtable}

\usepackage[b5paper]{geometry}
\begin{document}
	\mainmatter              
	\title{Radio pulsars --- two old questions \\
	(in blessed memory of Ya.N.Istomin)}
	\titlerunning{Radio pulsars}  
	%
	\author{Vasily Beskin\inst{1,2} }
	\authorrunning{Vasily Beskin} 
	%
	
	%
	\institute{P.N.Lebedev Physical Institute, 
	Moscow 119991, Russia,\\
		\email{beskin@lpi.ru}\\ 
		\and
		Moscow Institute of Physics and Technology, 
		Dolgoprudny 141701,
		 Russia}
	
	\maketitle

\begin{abstract}
	  We draw attention to two unanswered questions in radio pulsars' theory. The first one concerns the inclination angles $\chi$ between magnetic and rotation axes. We show that the very existence of interpulse pulsars indicates that all possible angles are realized. The second one is the question of the break of the death line on the $P {\dot P}$-diagram. We show that this break can be easily explained by a very mechanism of secondary plasma generation.  
\end{abstract}

\section{Introduction}

More than 50 years have passed since the discovery of radio pulsars. However, we still do not know 
the mechanism of their coherent radio emission, e.g., the direction of evolution of the inclination 
angle $\chi$. Some questions, formulated in the early years, were not further discussed. Here we outline
 the solution to one of such forgotten questions connecting with the possible range of inclination angles.

The second question concerns the so-called death line, i.e., the line limiting from below the population 
of pulsars on the $P{\dot P}$-diagram that has a break at $P \approx 0.1$ s. As we show, this break can 
be easily explained by the well-known feature of secondary plasma This work is dedicated to the blessed
 memory of Ya.N. Istomin, as well as B.V. Komberg, with whom we discussed these issues back in the 70s.

\section{Inclination angle}

If we take almost any catalog of the inclination angles $\chi$, we find that these angles lie in a range of $0 < \chi < 90^{\circ}$. But this does not mean that there are no radio pulsars with the inclination angle $\chi$ in a range of $90^{\circ} < \chi < 180^{\circ}$. This is simply because most of the existing methods for determining this angle do not distinguish between $\chi$ and $180^{\circ}$--$\chi$.

On the other hand, acute and obtuse angles between the axes correspond to two principally different physical conditions because for $\chi < 90^{\circ}$, the longitudinal current flowing along open magnetic field lines (with its sign determined by the sign of the~\citet{GJ} charge density $\rho_{\rm GJ} = -{\bf \Omega B}/2\pi c$) requires the outflow of negative charges from the star surface. By contrast, for the angles $\chi > 90^{\circ}$, the surface should eject positive ones. Hence, in models of plasma generation, in which the ejection of particles from the surface plays a significant role, these are two fundamentally different cases.
Recall that in~\citet{RS} model, particle creation is insensitive to axis orientation.

\begin{figure}
		\center{\includegraphics[width=0.9\linewidth]{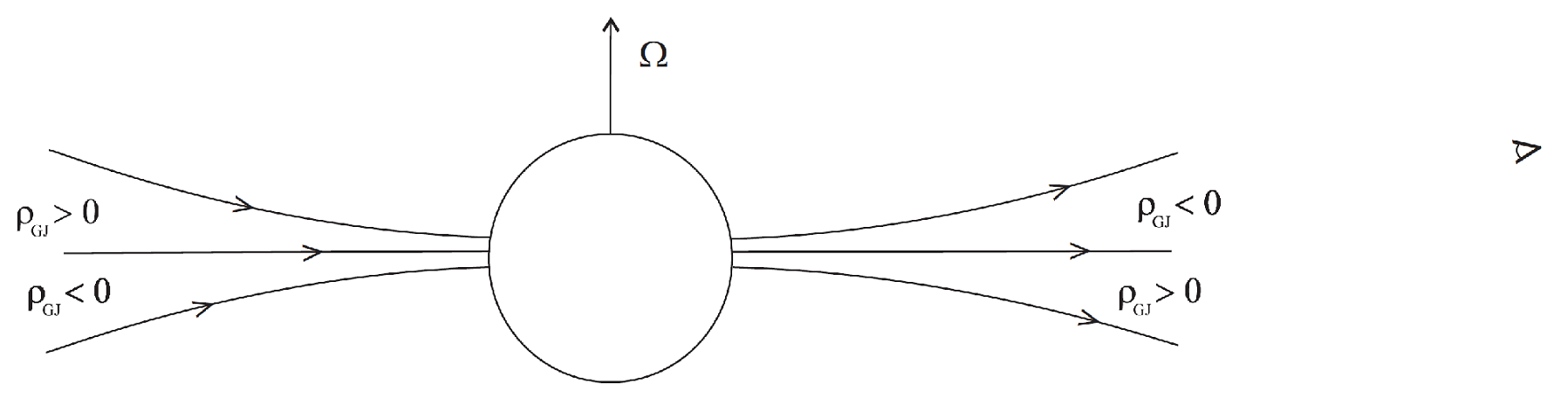}  }
	\caption{Observer located not on the equator sees the radio emission from two poles of the orthogonal rotator corresponding to two different signs of the charge density.}
\label{fig1}	
\end{figure}

At first glance, orthogonal pulsars give an immediate answer to this question. As shown in Fig.~\ref{fig1}, for $\chi = 90^{\circ}$, an observer
sees the regions of radio emission corresponding to different signs of the Goldreich-Julian charge density. However, in the case where the inclination angle $\chi$ differs from $90^{\circ}$, the line of sight can intersect the radiation patterns from two magnetic poles in the areas corresponding to the same charge density. Hence, a more detailed study is required.

We analyzed eight orthogonal interpulse pulsars presented by~\citet{SiMi19}, for which inclination angles $\chi$, the angles of closest approach of the line of sight to the magnetic axis $\beta$, and the emission heights $h$ were determined. This allowed us to estimate the inclination angles of magnetic field lines to the rotation axis $\delta$ in the plasma generation regions: \mbox{$\delta = \chi + \beta(R/h)^{1/2}$}. Here $R$ is the star radius. As shown from Table~\ref{tab}, for most pulsars, the angles $\delta$ for the main pulse and the interpulse correspond to different signs of the charge density. The uncertainty for two pulsars is apparently associated with a significant difference between the values of $h_{\rm MP}$ and $h_{\rm IP}$ (the accuracy of which is low). For other pulsars, our conclusion depends only slightly on the values of $h$. Thus, one can conclude that plasma ejection from the surface of a neutron star does not affect the process of particle production in the polar regions.

\begin{table}
	\caption{Interpulse pulsars data taken from~\citet{SiMi19}. All the angles are measured in degrees, and the radiation heights $h$ are in km.}
	\begin{center}
		\begin{tabular}{cccrcrrrrr}
			\hline
		\hspace{0.3cm}	PSR\hspace{0.3cm}	& \hspace{0.3cm}$P(s)$ \hspace{0.3cm}    &  \hspace{0.2cm}$\chi_{\rm MP}$  \hspace{0.2cm}  & \hspace{0.cm}$\beta_{\rm MP}$  &
        \hspace{0.4cm}$\chi_{\rm IP}$  \hspace{0.2cm}  & \hspace{0.2cm}$\beta_{\rm IP}$  &
		\hspace{0.4cm}$h_{\rm MP}$ &
		\hspace{0.4cm}$h_{\rm IP}$ &
		\hspace{0.4cm}$\delta_{\rm MP}$ &
		\hspace{0.4cm}$\delta_{\rm IP}$\\
			\hline
		 	J0627+0705   & 0.48 & 94.0 & $-$8.6 & 86.0 & $-$0.6 & 380 & 160 & 93 & 86 \\
		 	J0905$-$5157 & 0.35 & 90.4 & $-$3.6 & 89.5 &   $-$2.8 & 970 &  40 & 90 & 88 \\
		 	J0908$-$4913 & 0.11 & 83.6 &  7.4   & 96.4 &   $-$5.4 &  60 & 70  & 87 & 94 \\
		 	J1549$-$4848 & 0.29 & 87.3 &  2.3   & 92.7 &   $-$3.2 & 130 & 110 & 88 & 92 \\
		 	J1722$-$3712 & 0.24 & 87.4 & $-$5.6 & 92.6 &   $-$10.9& 110 & 300 & 86 & 91 \\
		 	J1739$-$2903 & 0.32 & 94.6 & $-$2.2 & 85.4 &    7.1    & 240 & 200 & 94 & 87 \\
		 	J1828$-$1101 & 0.07 & 82.7 &  7.3   & 97.3 &   $-$7.3 & 35  & 45  & 87 & 94 \\
		 	J1935+2025   & 0.08 & 93   & $-$13  & 87   &     $-$7   & 200 & 30  & 90 & 83 \\[2pt]
			\hline
		\end{tabular}
	\end{center}
	\label{tab}
\end{table}

\section{Death line}

\begin{figure}[!ht]
	\begin{minipage}{0.45\linewidth}
		\center{\includegraphics[width=1\linewidth]{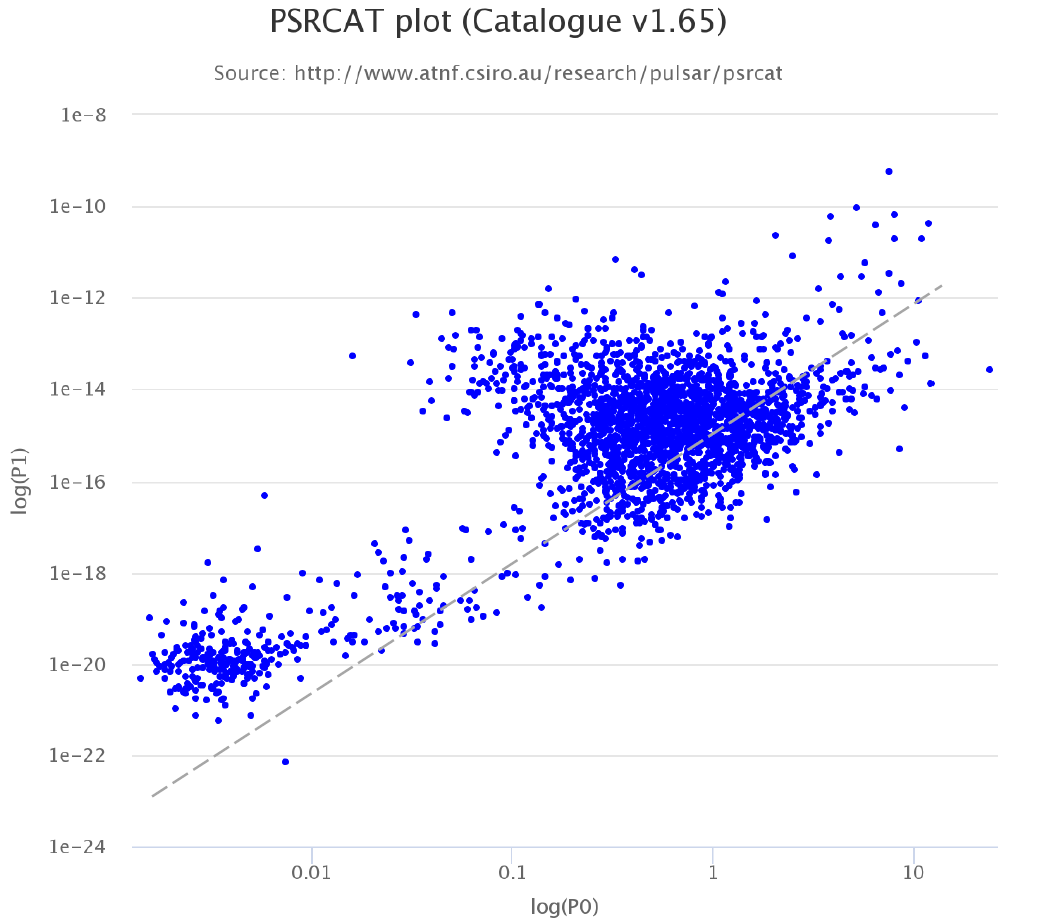} }
	\end{minipage}
	\hfill
	\begin{minipage}{0.45\linewidth}
		\center{\includegraphics[width=1\linewidth]{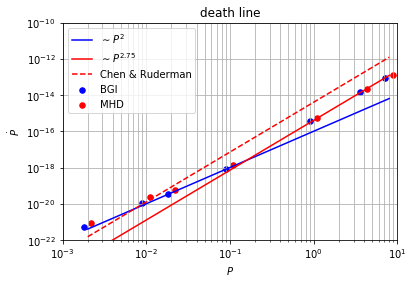} }
	\end{minipage}
 	\caption{$P{\dot P}$-diagram taken from the ATNF catalog~\citep{ATNF} (left) and the death line obtained for two models of pulsar braking (right). The dashed lines correspond to the ''classical'' death line presented by~\citet{ChRu93}.
	}
\label{fig2}	
\end{figure}

At the time of this writing, 3177 pulsars was already discovered~\citep{ATNF}. This rather rich statistics clearly shows that the line limiting from below the population of pulsars on the $P{\dot P}$-diagram has a break at $P \approx 0.1$ s (see Fig.~\ref{fig2}, left panel). Here we show that this break can be easily explained by the well-known feature of secondary plasma production.

Indeed, for pulsars with periods $P < 0.1$ s, the radiation reaction becomes significant, so the energy of primary particles does not reach the values dictated by the potential drop (see, e.g.,~\citealt{Jones2021}). Recently, we carried out a detailed study~\citep{BP2021} that took into account many still neglected phenomena, such as the role of high-energy $\gamma$-quanta in the synchrotron radiation spectrum, the possible wide distribution in moments of inertia, and, naturally, the effects of general relativity. In particular, the conclusion about the role of the radiation reaction for fast pulsars was confirmed.

A detailed analysis of this issue will be presented in a separate work. In the same article, we only show that even for standard values of the polar cap size and the parameters of a neutron star (and for dipole magnetic field!), taking into account the above effects significantly changes the position of the death line.

The right panel in Fig. 2 shows the position of the death line obtained from the condition of the absence of particle production for two models of pulsar braking (see~\citealt{BP2021} for more detail). The dashed lines correspond to the classic death line~\citep{ChRu93}, for which, by the way, unrealistic magneto-dipole losses were assumed. As we can see, regardless of the model of pulsar evolution, for $P > 0.1$ s, the death line locates one order of magnitude lower than the classical one. On the other hand, at $P < 0.1$ s, the slope becomes noticeably flatter (it corresponds to proportionality ${\dot P} \propto P^2$). Since our death line is still above many pulsars, a more detailed consideration is required, which we plan to carry out in the very near future.



I thank A.~Philippov and A.~Timokhin for the useful
discussions. This work was supported by Russian Foundation for Basic Research, grant 20-02-00469.

\bibliographystyle{aa}
\bibliography{Beskin_PSR}

\end{document}